\documentclass[pra,final,floatfix,twocolumn,showpacs]{revtex4-1}

\usepackage{graphicx}
\usepackage{dcolumn}
\usepackage{bm}
\usepackage{amssymb}
\usepackage{natbib}
\usepackage{amsmath}
\usepackage{color}
\usepackage{datetime}
\usepackage{footnote}
\usepackage{bbold}
\usepackage[normalem]{ulem}
\usepackage{dcolumn}
\usepackage{ragged2e}
\usepackage{xfrac}
\usepackage{sidecap}
\usepackage{setspace}
\usepackage{multirow}
\usepackage{xcolor}
\usepackage{booktabs}

\newcommand{\ket}[1]{\ensuremath{|#1\rangle}}

\newcommand{\gradi}[0]{^{\circ}}

\newcommand{\be}{\begin{equation}}
\newcommand{\ee}{\end{equation}}

\newcommand{\1}{{\mathbbm{1}}}

\newcommand{\nocontentsline}[3]{}
\newcommand{\tocless}[2]{\bgroup\let\addcontentsline=\nocontentsline#1{#2}\egroup}

\newif\iffigs
\figstrue

\begin{document}

\title{Experimental Perfect Quantum State Transfer}

\author{Robert J. Chapman$^{1,2}$}

\author{Matteo Santandrea$^{3,4}$}

\author{Zixin Huang$^{1,2}$}

\author{Giacomo Corrielli$^{3,4}$}

\author{Andrea Crespi$^{3,4}$}

\author{Man-Hong Yung$^{5}$}

\author{Roberto Osellame$^{3,4}$}

\author{Alberto Peruzzo$^{1,2}$}
\email{alberto.peruzzo@rmit.edu.au}

\affiliation{
$^{1}$Quantum Photonics Laboratory, School of Engineering, RMIT University, Melbourne, Australia\\
$^{2}$School of Physics, The University of Sydney, Sydney, New South Wales 2006, Australia\\
$^{3}$Istituto di Fotonica e Nanotecnologie, Consiglio Nazionale delle Ricerche, Piazza Leonardo da Vinci 32, I-20133 Milano, Italy\\
$^{4}$Dipartimento di Fisica, Politecnico di Milano, Piazza Leonardo da Vinci 32, I-20133 Milano, Italy\\
$^{5}$Department of Physics, South University of Science and Technology of China, Shenzhen, 518055, P. R. China
}

\begin{abstract}
The transfer of data is a fundamental task in information systems. Microprocessors contain dedicated data buses that transmit bits across different locations and implement sophisticated routing protocols. Transferring quantum information with high fidelity is a challenging task, due to the intrinsic fragility of quantum states. We report on the implementation of the perfect state transfer protocol applied to a photonic qubit entangled with another qubit at a different location. On a single device we perform three routing procedures on entangled states with an average fidelity of 97.1\%. Our protocol extends the regular perfect state transfer by maintaining quantum information encoded in the polarisation state of the photonic qubit. Our results demonstrate the key principle of perfect state transfer, opening a route toward data transfer for quantum computing systems.
\end{abstract}

\maketitle


Transferring quantum information between distant lattice sites without disrupting the encoded information en route is crucial for future quantum technologies \cite{nielsen_quantum_????,bose_quantum_2003,christandl_perfect_2004,divincenzo_physical_2000}. Cascaded SWAP operations between neighbouring sites enable quantum information transfer, however, this method is intrinsically weak as individual errors accumulate after each operation, leading to a heavy decrease in transfer success. A more robust scheme designed to minimize dynamic control is therefore required.

\begin{figure}[t]
\centering
\includegraphics[width=1.0\linewidth]{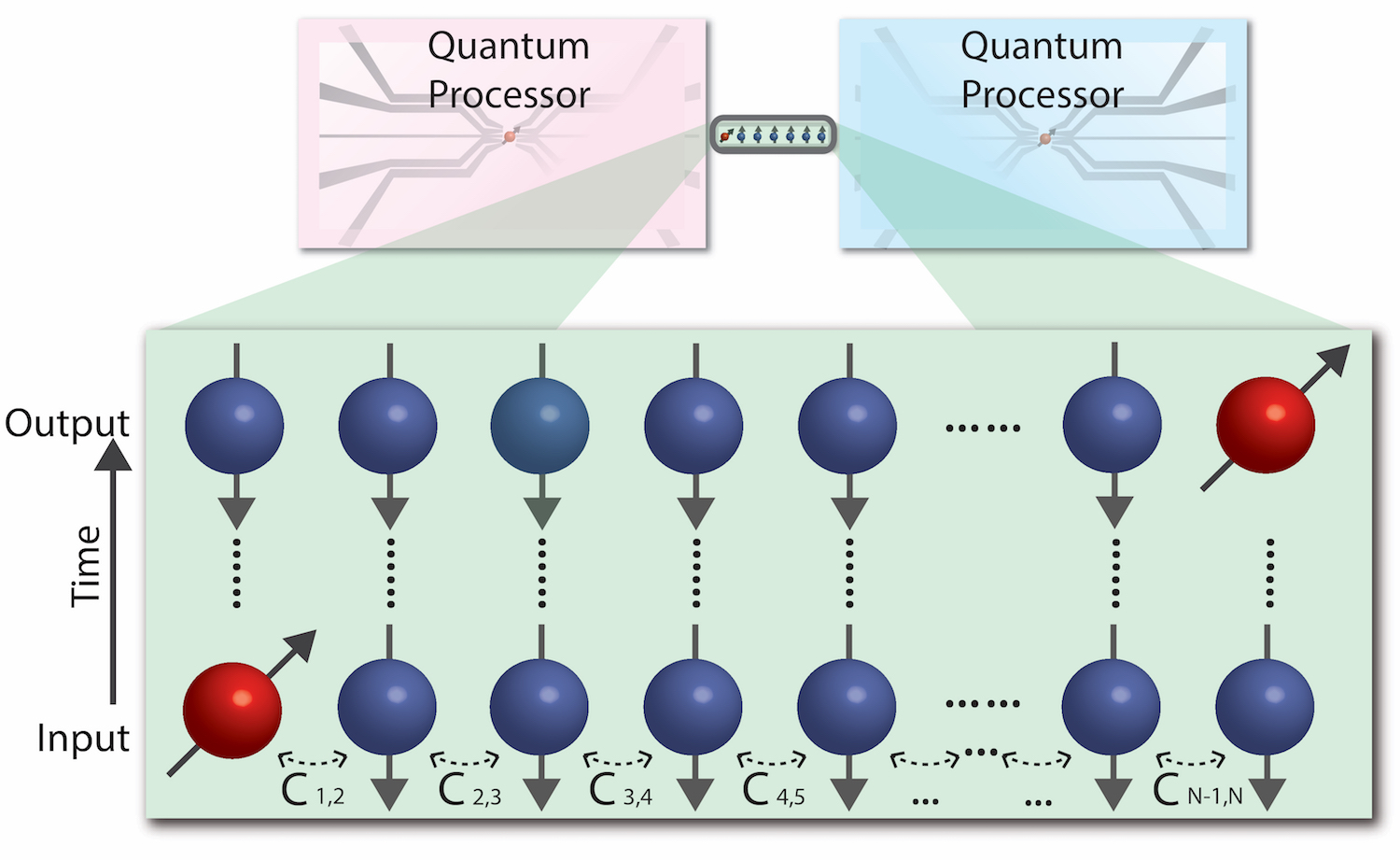}
\caption{\textbf{Illustration of a one-dimensional perfect state transfer lattice connecting two quantum processors.} The state at the first site is recovered at the final site after a specific time.}
\label{figure:pst_chain}
\end{figure}

The perfect state transfer (PST) protocol utilises an engineered but fixed coupled lattice. Quantum states are transferred between sites through Hamiltonian evolution for a specified time \cite{bose_quantum_2003,christandl_perfect_2004}. For a one-dimensional system with $N$ sites, the state at site $n$ is transferred to site $N-n+1$ with 100\% probability \cite{christandl_perfect_2005}. PST can be performed on any quantum computing architecture where coupling between sites can be engineered, such as ion traps \cite{kielpinski_architecture_2002} and quantum dots \cite{loss_quantum_1998}. Figure \ref{figure:pst_chain} presents an illustration of the PST protocol. The encoded quantum state, initially at the first site, is recovered at the final site after a specific time. Aside from qubit relocation, the PST framework can be applied to entangled W-state preparation \cite{kay_perfect_2010}, state amplification \cite{kay_unifying_2007} and even quantum computation \cite{kay_computational_2008}.

To date, most research on PST has been theoretical \cite{bose_quantum_2003,christandl_perfect_2004,yung_perfect_2005,christandl_perfect_2005,yung_quantum_2006,kay_perfect_2006,
bose_quantum_2007,kay_unifying_2007,kay_computational_2008,kay_perfect_2010,perez-leija_perfect_2013} with experiments \cite{bellec_faithful_2012,perez-leija_coherent_2013} being limited to demonstrations where no quantum information is transferred and do not incorporate entanglement, often considered the defining feature of quantum mechanics \cite{einstein_can_1935}. We present the implementation of a protocol that extends PST for relocating a polarisation encoded photonic qubit across a one-dimensional lattice, realised as an array of 11 evanescently coupled waveguides \cite{perets_realization_2008,rai_transport_2008,peruzzo_quantum_2010}. We show that the entanglement between a photon propagating through the PST waveguide array and another photon at a different location is preserved.

\begin{figure}[t]
\centering
\includegraphics[width=1.0\linewidth]{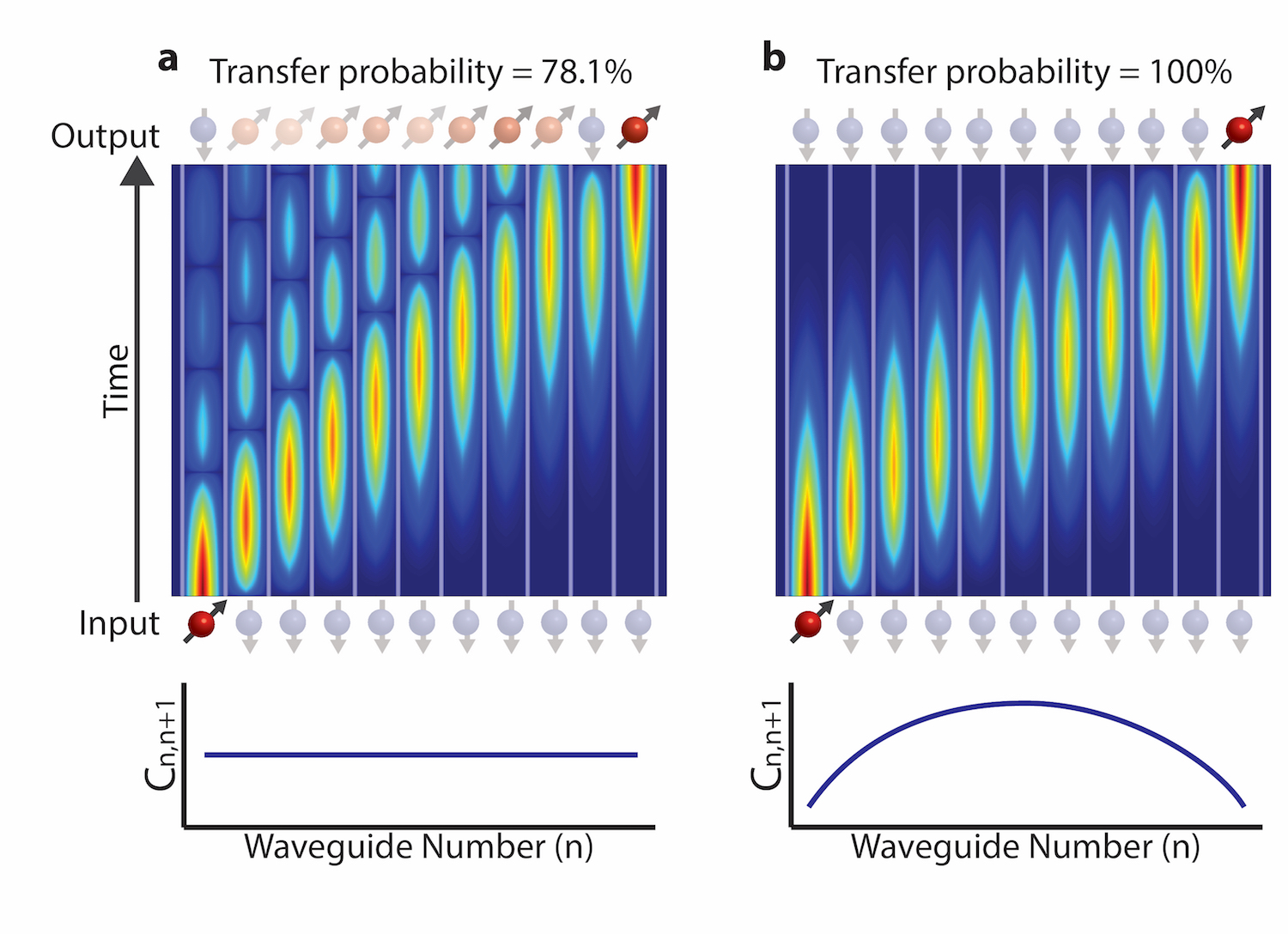}
\caption{\textbf{Propagation simulations with different coupling coefficient spectra.} \textbf{a}, A photon is injected into the first waveguide of an 11 waveguide array with the Hamiltonian in equation (\ref{equation:hamiltonian}) and a uniform coupling coefficient spectrum. With the constraint that reflections off boundaries are not allowed, we calculate a maximum probability of transferring the photon to waveguide 11 of 78.1\% \cite{bose_quantum_2003}. \textbf{b}, A photon is injected into the first waveguide of an 11 waveguide array, this time with the coupling coefficient spectrum of equation (\ref{equation:coupling}). After evolution for a pre-determined time, the photon is received at waveguide 11 with 100\% probability \cite{christandl_perfect_2004}.}
\label{figure:pst_array}
\end{figure}

\begin{figure*}
\centering
\includegraphics[width=1.0\linewidth]{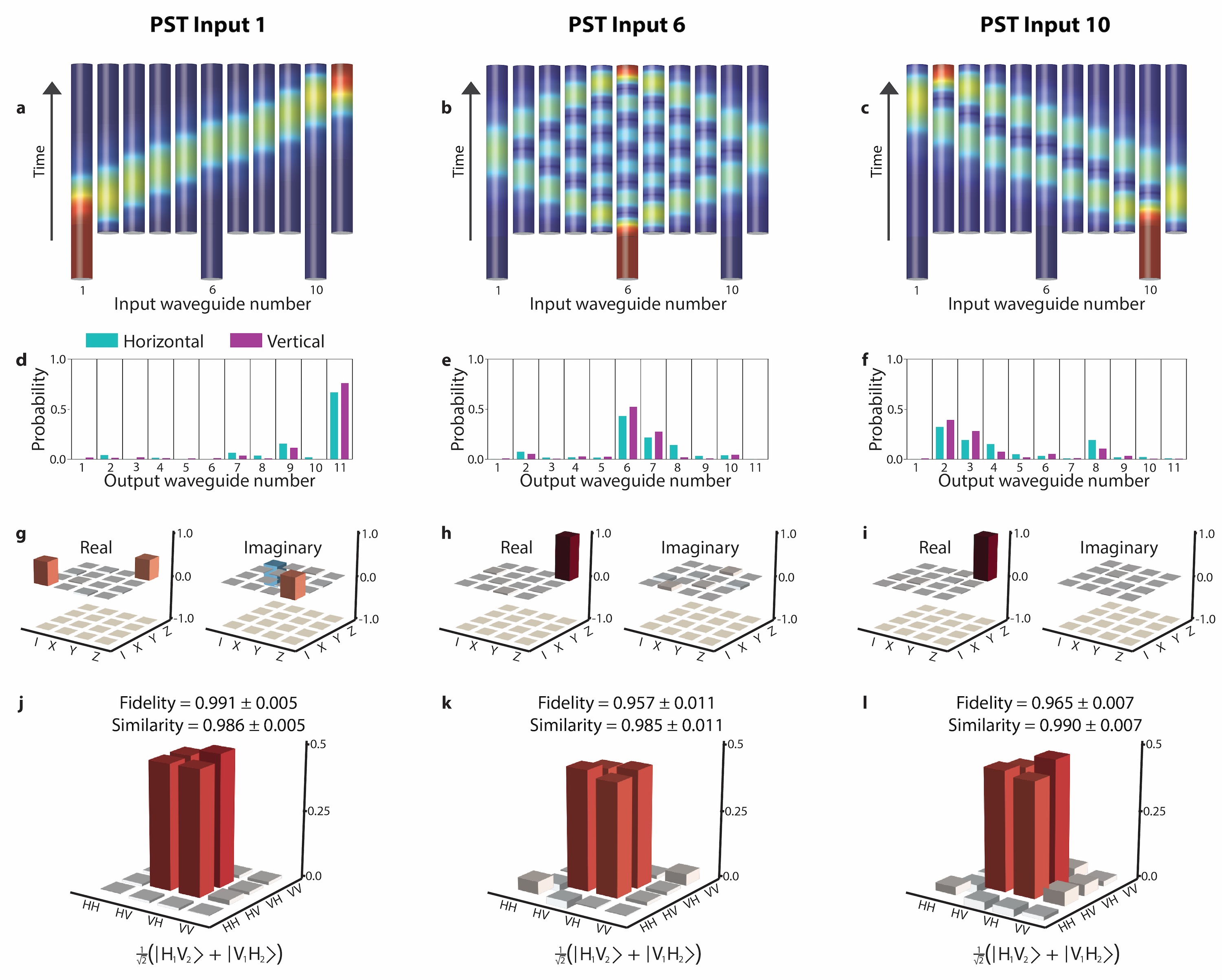}
\caption{\textbf{Perfect state transfer probability distribution characterisation, quantum process tomography and entangled state transfer.} \textbf{a-c}, Propagation simulations showing the device implementation to enable specific waveguide input. \textbf{d-f}, Output probability distributions for each input of the PST array for horizontally and vertically polarised laser light. \textbf{g-i}, Quantum process matrix for each transfer in the PST array measured with single photon quantum process tomography. \textbf{j-l}, Two photon quantum state tomography is performed after photon 1 of the polarisation entangled Bell state $\tfrac{1}{\sqrt{2}}(\ket{\mathrm{H}_1\mathrm{V}_2} + \ket{\mathrm{V}_1\mathrm{H}_2})$ has been relocated. Results have had the small imaginary components removed for brevity.}
\label{figure:results}
\end{figure*}

The Hamiltonian for our system in the nearest-neighbour approximation is given by the tight binding formalism

\begin{equation}
H = \sum_{\sigma \in \{\mathrm{H,V}\}} \sum_{n=1}^{N-1}C_{n,n+1}\big(\hat{a}_{n+1,\sigma}^{\dagger}\hat{a}_{n,\sigma} + \hat{a}_{n,\sigma}^{\dagger}\hat{a}_{n+1,\sigma}\big),
\label{equation:hamiltonian}
\end{equation}•

\noindent where $C_{n,n+1}$ is the coupling coefficient between waveguides $n$ and $n+1$, and $\hat{a}_{n,\sigma}$ ($\hat{a}_{n,\sigma}^{\dagger}$) is the annihilation (creation) operator applied to waveguide $n$ and polarisation $\sigma$. Hamiltonian evolution of a state $\ket{\psi_0}$ for a time $t$ is calculated via the Schr\"{o}dinger equation giving the final state $\ket{\psi(t)} = \exp(-iHt)\ket{\psi_0}$ \cite{bromberg_quantum_2009}. Equation (\ref{equation:hamiltonian}) is constructed of independent tight-binding Hamiltonians acting on each orthogonal polarisation. This requires there to be no cross-talk terms $\hat{a}_{n,\mathrm{H}}^{\dagger}\hat{a}_{m,\mathrm{V}}$ or $\hat{a}_{n,\mathrm{V}}^{\dagger}\hat{a}_{m,\mathrm{H}}\;\forall\; m,n$. The spectrum of coupling coefficients $C_{n,n+1}$ is crucial for successful PST. Evolution of this Hamiltonian with a uniform coupling coefficient spectrum, equivalent to equally spaced waveguides, is not sufficient for PST with over three lattice sites as simulated in figure \ref{figure:pst_array}a. PST requires the coupling coefficient spectrum to follow the function 

\begin{equation}
C_{n,n+1} = C_0\sqrt{n(N-n)},
\label{equation:coupling}
\end{equation}

\noindent where $C_0$ is a constant, $N$ is the total number of lattice sites and evolution is for a specific time $t_{\mathrm{PST}} = \tfrac{\pi}{2C_0}$ \cite{christandl_perfect_2004}. This enables arbitrary length PST as simulated in figure \ref{figure:pst_array}b for 11 sites. The coupling coefficient spectrum for each polarisation must be equal and follow equation (\ref{equation:coupling}) for the qubit to be faithfully relocated and the polarisation encoded quantum information to be preserved. The distance between waveguides dictates the coupling coefficient, however, for planar systems, the coupling coefficient of each polarisation will in general be unequal due to the waveguide birefringence. In order to achieve equal coupling between polarisations, the waveguide array is fabricated along a tilted plane in the substrate \cite{sansoni_two-particle_2012}. This is made possible by the unique three-dimensional capabilities of the femtosecond laser writing technique (see Supplementary Section \ref{supplementary:fabrication} for further fabrication and device details). We inject photons into waveguides 1, 6 and 10 of the array which after time $t_{\mathrm{PST}}$ transfer to waveguides 11, 6 and 2, respectively. Figures \ref{figure:results}a-c present propagation simulations for each transfer. Input waveguides extend to the end of the device to allow selective injection.
\\

In order to characterise the coupling coefficient spectra, we inject horizontally and vertically polarized laser light at 808 nm into each input waveguide. Laser light is more robust to noise than single photons and we can monitor the output with a CCD camera to fast gather results. Using laser light at the same wavelength as our single photons will give an output intensity distribution equivalent to the output probability distribution for detecting single photons \cite{perets_realization_2008}. Ideally light injected into waveguide $n$ will output the device only in waveguide $N-n+1$, however, this assumes an approximate model of nearest-neighbour coupling only. Taking into account coupling between further separated waveguides reduces the transfer probability. This decrease is greater for light injected closer to the centre of the array. Figures \ref{figure:results}d-f present our measured output probability distribution for horizontally ($P_n^\mathrm{H}$) and vertically ($P_n^\mathrm{V}$) polarised laser light injected into each input waveguide, where $n$ is the output waveguide number. Fidelity between the probability distributions for each polarisation is given by $F_{\mathrm{distribution}} = \big(\sum_n\sqrt{P_n^\mathrm{H}P_n^\mathrm{V}}\big)^2$. This fidelity is closely related to how similar the two coupling coefficient spectra are. We measure an average probability distribution fidelity for all transfers of $0.949 \pm 0.007$ (see Supplementary Section \ref{supplementary:fabrication} for all fidelity values). We encode quantum information in the polarisation state of the photon and are interested in reliably relocating this qubit. We use a single optical fibre to capture photons from the designed output waveguide which, in all cases, is the waveguide with the greatest output probability.
\\

We perform quantum process tomography to understand the operation performed on the single photon polarisation state during each PST transfer. We inject single photon states $\ket{\psi_{\mathrm{in}}} = (\alpha\hat{a}_{\mathrm{S},\mathrm{H}}^{\dagger}+\beta\hat{a}_{\mathrm{S},\mathrm{V}}^{\dagger})\ket{0}$ into each input waveguide $\mathrm{S} \in \{1,6,10\}$, where $\alpha$ ($\beta$) is the probability amplitude of the horizontal (vertical) component of the photon. From quantum process tomography on the output polarisation states, we can generate a process matrix $\chi_{\mathrm{pol}}$ for each transfer \cite{nielsen_quantum_????,obrien_quantum_2004}. We aim to perform identity so the quantum information encoded in the polarisation can be recovered after relocation. We measure a polarisation phase shift associated with each transfer. This phase shift can be compensated for with a local polarisation rotation applied before injection. Figures \ref{figure:results}g-i present our measured process matrix for each transfer. Across all transfers we demonstrate an average fidelity of the polarisation process including compensation to identity of $0.982 \pm 0.003$ (see Supplementary Section \ref{supplementary:qpt} for details of the compensation scheme and all fidelities). Process fidelity is calculated as $F_{\mathrm{process}} = Tr\{\chi_{\mathbb{1}}\chi_{\mathrm{pol+comp}}\}$ \cite{gilchrist_distance_2005} where $\chi_{\mathbb{1}}$ is the process matrix for the identity operation and $\chi_{\mathrm{pol+comp}}$ is the combined polarisation operation and compensation process matrix.

Ideally the output state for each transfer is $\ket{\psi_{\mathrm{out}}} = (\alpha\hat{a}_{\mathrm{T},\mathrm{H}}^{\dagger}+\beta\hat{a}_{\mathrm{T},\mathrm{V}}^{\dagger})\ket{0}$ where $\mathrm{T} \in \{11,6,2\}$ and the probability amplitude of each polarisation component remains equal to the input state. Our high fidelity measurements on single photon relocation demonstrate that we can route a polarisation encoded photonic qubit across our device and faithfully recover the quantum information.
\\

\begin{figure*}
\centering
\includegraphics[width=1.0\linewidth]{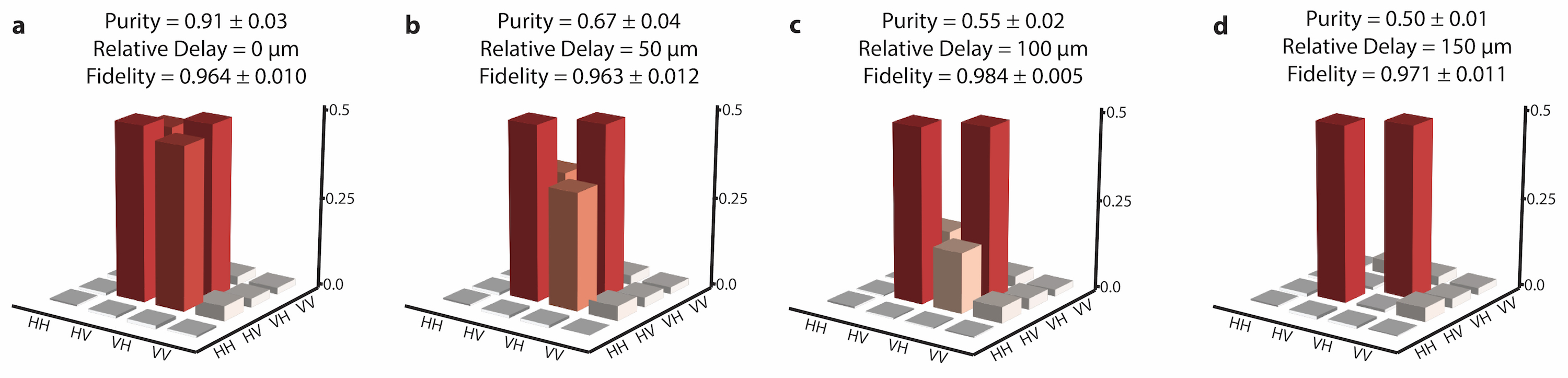}
\caption{\textbf{Perfect state transfer of entangled states with varying purity.} Photon 1 of the state $\tfrac{1}{\sqrt{2}}(\ket{\mathrm{H}_1\mathrm{V}_2} + \ket{\mathrm{V}_1\mathrm{H}_2})$ is injected into waveguide 1 of the PST array. A delay is applied to the vertical component in order to control the purity of the state. \textbf{a}, Relative delay of 0 $\mu\mathrm{m}$, \textbf{b}, 50 $\mu\mathrm{m}$, \textbf{c}, 100 $\mu\mathrm{m}$ and \textbf{d}, 150 $\mu\mathrm{m}$. Results have had the small imaginary components removed for brevity.}
\label{figure:decohered}
\end{figure*}

Entanglement is likely to be a defining feature of quantum computing and preserving entanglement is therefore critical to the success of any qubit relocation protocol. We prepare the Bell state $\tfrac{1}{\sqrt{2}}(\ket{\mathrm{H}_1\mathrm{V}_2} + \ket{\mathrm{V}_1\mathrm{H}_2})$ using spontaneous parametric down conversion process \cite{matthews_observing_2013} (see Supplementary Section \ref{supplementary:experiment} for details). Photon 1 is injected into the waveguide array while photon 2 propagates through polarisation maintaining fibre (PMF). In terms of waveguide occupancy, our input state is $\ket{\Psi_\mathrm{in}} = \tfrac{1}{\sqrt{2}}(\hat{a}_{\mathrm{S},\mathrm{H}}^{\dagger}\hat{a}_{0,\mathrm{V}}^{\dagger} + \hat{a}_{\mathrm{S},\mathrm{V}}^{\dagger}\hat{a}_{0,\mathrm{H}}^{\dagger})\ket{00}$ for each input waveguide $\mathrm{S} \in \{1,6,10\}$, where $\hat{a}_{0,\mathrm{\sigma}}^{\dagger}$ denotes the creation operator acting on polarisation $\sigma$ in PMF. Full two qubit polarisation tomography \cite{james_measurement_2001} is performed on the output and the fidelity calculated as $F_{\mathrm{quantum}} = \big(Tr\big\{\sqrt{\sqrt{\rho_1}\rho_2\sqrt{\rho_1}}\big\}\big)^2$ \cite{jozsa_fidelity_1994} where $\rho_1$ is the density matrix after the PST protocol has been applied and $\rho_2$ is the density matrix after propagation through a reference straight waveguide. After all qubit relocations we measure an average polarisation state fidelity of $0.971 \pm 0.014$. Using results from quantum process tomography, we can predict the state after PST and calculated the similarity; the fidelity between our measured state and predicted state. We calculate an average similarity of $0.987 \pm 0.014$ across all transfers (see Supplementary Section \ref{supplementary:fidelities} for all fidelities and similarities). Figures \ref{figure:results}j-l present our measured density matrix after each entangled state transfer.

Ideally, the output state for each transfer is $\ket{\Psi_\mathrm{out}} = \tfrac{1}{\sqrt{2}}(\hat{a}_{\mathrm{T},\mathrm{H}}^{\dagger}\hat{a}_{0,\mathrm{V}}^{\dagger} + \hat{a}_{\mathrm{T},\mathrm{V}}^{\dagger}\hat{a}_{0,\mathrm{H}}^{\dagger})\ket{00}$ where $\mathrm{T} \in \{11,6,2\}$. With high fidelity the probability amplitude of each component is preserved and the state remains almost pure. This result demonstrates that with our device we can relocate a polarisation qubit between distant sites and preserve entanglement with another qubit at a different location. In principle our device could route qubits from any waveguide $n$ to waveguide $N-n+1$. Quantum error correction protocols require sophisticated interconnection in order to access individual qubits for control and measurement within large, highly entangled surface code geometries \cite{devitt_quantum_2013}. PST is a clear gateway towards accessing qubits in such systems without disrupting quantum states and entanglement throughout the surface code.
\\

Decoherence has applications in quantum simulation to emulate systems in nature \cite{lloyd_universal_1996} and it is therefore important to note that this approach for relocating quantum information can be applied to states of any purity \cite{christandl_perfect_2004}. We prepare decohered states by introducing a time delay between the horizontal and vertical components of the polarisation qubit. The purity of the state can be calculated as the convolution of the horizontal and vertical components with a time delay $\tau$
\begin{equation}
\mathrm{Purity}(\tau) \equiv \int_{-\infty}^{\infty} \mathrm{H}(t)\mathrm{V}(\tau-t)dt,
\end{equation}
\noindent where $\tau$ is controlled by altering the path length of the vertical component of the state. H (V) is the horizontal (vertical) component of the photon. Figures \ref{figure:decohered}a-d present density matrices for PST from waveguide 1 to waveguide 11 applied to entangled states of varying purity. The injected states are recovered with an average fidelity of $0.971 \pm 0.019$ and average similarity of $0.978 \pm 0.019$ (see Supplementary Section \ref{supplementary:fidelities} for all values).
\\

We have proposed and experimentally demonstrated a protocol for relocating a photonic qubit across 11 discrete sites maintaining the quantum state with high fidelity and preserving entanglement with another qubit at a different location. We can aim to improve our fidelity by reducing next-nearest-neighbour coupling via further separating the waveguides and having a longer device. This would increase the contrast between nearest- and next-nearest-neighbour coupling to better fit the Hamiltonian in equation (\ref{equation:hamiltonian}). A bi-product of longer devices, however, is an increase in propagation loss. Depth-dependent spherical aberrations in the laser irradiation process may also affect the homogeneity of the three-dimensional waveguide array. Additional optics in the laser writing setup could be employed to reduce this effect. Protocols for relocating quantum information across discrete sites are essential for future quantum technologies. Our protocol builds on the perfect state transfer with extension to include an additional degree of freedom for encoding quantum information. This demonstration opens pathways toward faithful quantum state relocation in quantum computing systems.

\newpage

\bibliographystyle{apsrev4-1}


%

\clearpage

\section*{Supplementary Information}

\setcounter{figure}{0}
\makeatletter 
\renewcommand{\thefigure}{S\@arabic\c@figure}
\makeatother

\setcounter{equation}{0}
\makeatletter 
\renewcommand{\theequation}{S\@arabic\c@equation}
\makeatother

\setcounter{table}{0}
\makeatletter 
\renewcommand{\thetable}{S\@arabic\c@table}
\makeatother

\section{Technical Fabrication Details} \label{supplementary:fabrication}

\paragraph*{Waveguide fabrication} 
Integrated photonic waveguides with single mode propagation at 808 nm are fabricated by focusing femtosecond laser pulses with an energy of 300 $\mathrm{nJ/pulse}$, at the repetition rate of 1 $\mathrm{MHz}$, in the bulk of a borosilicate substrate (Eagle2000, Corning) by means of a 20$\times$ microscope objective (NA = 0.45, Achroplan) and translating the sample at the constant speed of 40 $\mathrm{mm/s}$.\\
The resulting waveguides exhibit relatively low propagation losses (0.8 $\mathrm{dB/cm}$) and elliptical guided mode ($\mathrm{1/e^2}$ diameters measured as 9.4 $\mathrm{\mu m}$ $\times$ 15.1 $\mathrm{\mu m}$).

\paragraph*{Design of the array}
The inter-waveguide distances are not uniform and are specially tailored in order to implement the correct couplings expressed by equation 2 in the main text, which allow the perfect state transfer protocol. The coupling between waveguides is a function of their separation $d$, according to the formula
\begin{equation}\label{DECAY}
C(d) = ae^{-bd},
\end{equation}
where $a$ and $b$ are constants whose values have been measured as $a=3.6\,mm^{-1}$ and $b=0.19\, \mathrm{\mu m}^{-1}$. The distance $d_n$ between waveguide $n$ and waveguide $n+1$ of the array can be parametrised as follows:
\begin{equation}
d_n = d_{min}+\frac{1}{b}\log\left[\frac{1}{2}\sqrt{\frac{N^2-1}{n\left(N-n\right)}}\right],
\end{equation}
where $d_{min}$ is a free fabrication parameter that represents the minimum distance in the array and N (odd for this equation) is the total number of waveguides. With this parametrisation, the state transfer distance $z_{PST}$ can be expressed as:
\begin{equation}\label{ZPST}
z_{PST}=\frac{\pi\sqrt{N^2-1}}{4C_{max}},
\end{equation}
where $C_{max}=C(d_{min})$. The choice of the values $N$ and $d_{min}$ must be done in order to minimise the array non-idealities such as propagation losses, array inhomogeneity and parasitic couplings between non adjacent sites. In our design we have chosen the values $N=11$ and $d_{min}=12\,\mathrm{\mu m}$. Consequently, we fabricated 16 arrays with different lengths, spanning around the theoretical refocusing distance $z_{PST}^{th}$ = 23 mm, ranging from 21.5 to 29 mm, and the best results were observed for an array length of 22.5 mm. In order to ensure that the inter-waveguide couplings for horizontally and vertically polarised light are the same, the arrays extend diagonally into the substrate, at an angle of $\approx 60\gradi$. The central waveguide of each array, corresponding to waveguide 6, is situated 170 $\mathrm{\mu m}$ below the sample top surface. Finally, in order to couple light selectively in a given waveguide of the array, only waveguides identified by label 1, 6 and 10 reach the input facet of the devices, as depicted in Figures \ref{figure:results}a-c of the main text. \\

\paragraph*{Classical Characterisation}
In order to characterise the performances of the fabricated structures, we injected laser light in each of the available input ports and we measured the corresponding near field intensity profiles $I_n^\sigma$ at the arrays output via a CCD camera and suitable imaging optics (label $n$ indicates the waveguide number and $\sigma$ the polarisation). We used laser light at 808 nm, which is very close to the wavelength employed in the single photon experiments. The output probability distribution $P_n^\sigma$ for polarisation $\sigma$ is then defined as:
\begin{equation}
P_n ^\sigma= \frac{I_n^\sigma}{\sum_n I_n^\sigma}.
\end{equation}
We repeated the experiment for both horizontally and vertically polarised light and the output probability distributions are shown in Figures \ref{figure:results}d-f of the main text. In order to quantify the polarisation insensitivity of our devices we introduce the polarisation fidelity parameter $F_{\mathrm{distribution}}$, calculated as:
\begin{equation}
F_{\mathrm{distribution}} = \frac{\left(\sum_n\sqrt{I_n^\mathrm{H} I_n^\mathrm{V}} \right)^2}{\sum_n I_n^\mathrm{H} \sum_n I_n^\mathrm{V}} = \left(\sum_n \sqrt{P_n^\mathrm{H} P_n^\mathrm{V}}\right)^2.
\end{equation}
Table S1 gives the values of the distribution fidelities corresponding to each input waveguide.

\begin{table}[h]
	\begin{tabular}{c  c}
    \hline \hline
\bf{Input waveguide}&\bf{Distribution Fidelity} \\ 

1 & $0.957\pm 0.005$ \\ 
6 &  $0.945 \pm 0.003$ \\
10 & $0.946 \pm 0.004$ \\ \hline
    \end{tabular}
\caption{Fidelity between output intensity profiles for horizontally and vertically polarised laser light injected into each input waveguide.}
\label{table:classical}
\end{table}

\section{Quantum Process Tomography} \label{supplementary:qpt}

Quantum process tomography (QPT) characterises an unknown quantum process by performing state tomography on a range of output states. We perform single qubit QPT using input states $\ket{H}$, $\ket{V}$, $\ket{D} = \tfrac{1}{\sqrt{2}}(\ket{H} + \ket{V})$ and $\ket{R} = \tfrac{1}{\sqrt{2}}(\ket{H} + i\ket{V})$. Each state is prepared through a straight waveguide, before switching to the PST array.

\begin{table}[h]
	\begin{tabular}{ c  c  c  c }
    \hline \hline
\bf{Input}&\bf{HWP}&\bf{Phase}&\bf{Fidelity to}  \\ 
\bf{Waveguide}&\bf{(degrees)}&\bf{(degrees)} & \bf{$\chi_{\mathbb{1}}$} \\ 
1 & 0.438 & -68.5 & $0.986\pm0.002$ \\ 
6 & 0.349 & 44.7 & $0.975\pm0.002$ \\
10 & 0.453 & 45.3 & $0.984\pm0.002$ \\ \hline
    \end{tabular}
\caption{Compensation parameters to achieve identity on the polarisation state after the PST protocol has been applied.}
\label{table:comp}
\end{table}

The output $\chi$ matrices are shown in figures \ref{figure:results}g-i of the main text. From these process matrices we can determine a compensation scheme in terms of an HWP rotation and polarisation phase rotation. This is the same scheme as the state preparation and so the state preparation and compensation can be combined. The waveplate and phase angles of the pre-compensation scheme are given in table \ref{table:comp}.

Process fidelity values describe how close the overall process (compensation and measured PST) is to identity on the polarisation. This is calculated as the trace distance between identity and the polarisation process $Tr\{\chi_{\mathbb{1}}\chi_{\mathrm{pol+comp}}\}$. The compensation scheme gives a very good fidelity on how close to identity the whole operation will be. The reason it is not 100\% is a result of some decoherence in the system, which cannot be compensated for with local unitary operations.

\section{Experimental Setup} \label{supplementary:experiment}

\paragraph*{Polarisation entangled source}

Horizontally polarised photon pairs at 807.5 nm are generated via type 1 spontaneous parametric down conversion in a 1 mm thick BiBO crystal, pumped by an 80 mW, 403.75 nm CW diode laser. Both photons are rotated into a diagonal state $\tfrac{1}{\sqrt{2}}(\ket{H} + \ket{V})$ by a half waveplate (HWP) with fast axis at 22.5$\gradi$ from vertical. One photon has a phase applied by two 45$\gradi$ quarter waveplates (QWP) either side of a HWP at $\theta\gradi$. The second photon has its diagonal state optimised with a polarising beam splitter (PBS) at $\sim45\gradi$. 

\begin{figure}[t]
\centering
\includegraphics[width=1.0\linewidth]{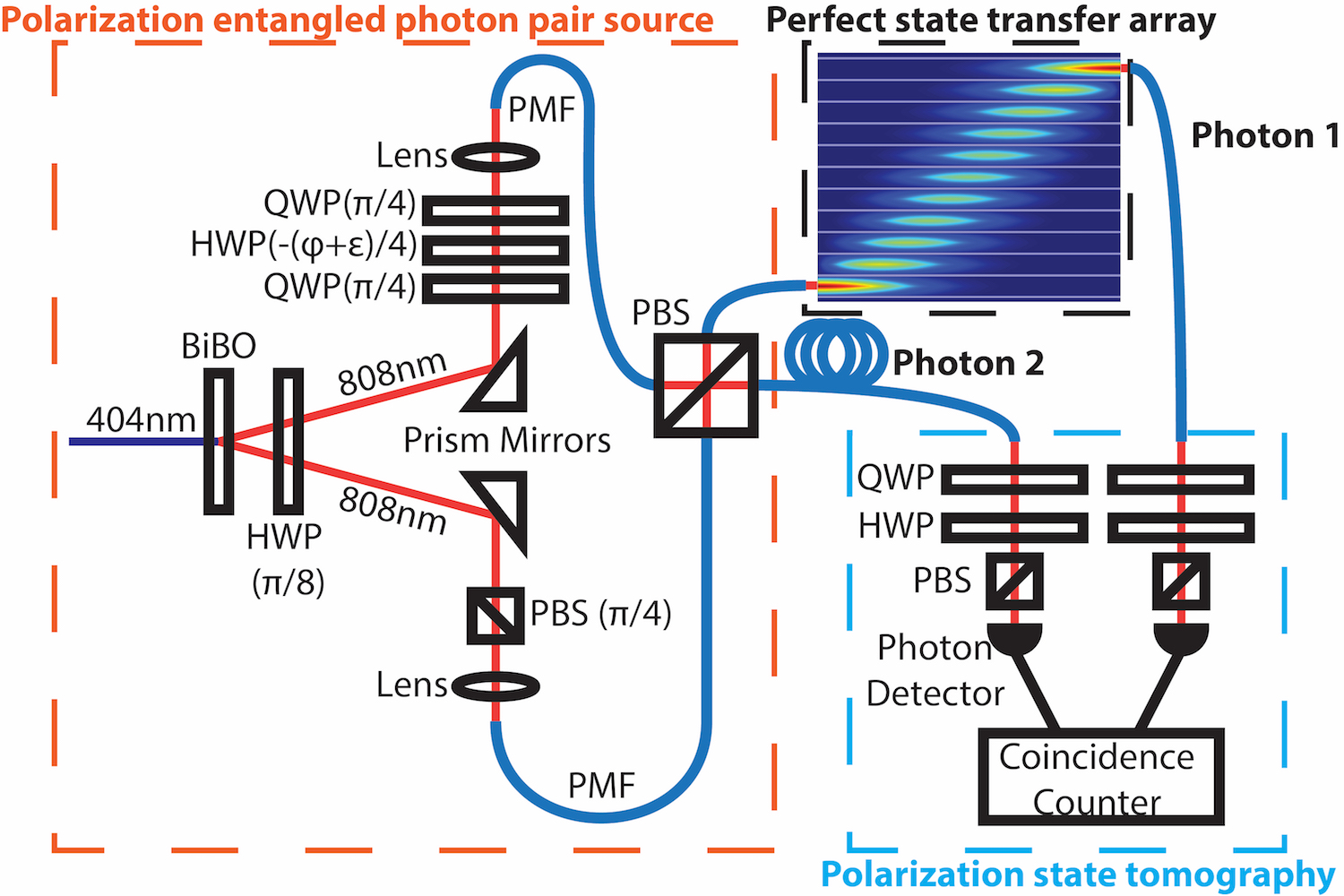}
\caption{Experimental setup. polarisation entangled photons are generated in free space before coupling into polarisation maintaining fibre (PMF). Photon 1 is injected into the perfect state transfer array while photon 2 travels through PMF. Full two qubit polarisation tomography is performed on the output.}
\label{figure:experiment}
\end{figure}

Both photons are collected in polarisation maintaining fibre (PMF) and are incident on both faces of a fiber pigtailed PBS. When measuring in the coincidence basis, this post-selects the entangled state $\tfrac{1}{\sqrt{2}}(\ket{H_1V_2} + e^{i\phi}\ket{V_1H_2})$ where $\phi = 4(\theta + \epsilon)$ and $\epsilon$ is the intrinsic phase applied by the whole system. The experimental setup is illustrated in figure \ref{figure:experiment}.

PMF is highly birefringent, resulting in full decoherence of the polarisation state after $\sim1$ m of fibre giving a mixed state. In order to maintain polarisation superposition over several meters of fiber we use 90$\gradi$ connections to ensure both polarisations propagate through equal proportions of fast and slow axis fibre. Slight length differences between fibres and temperature variations mean the whole system applies a residual phase $\epsilon$ to the state, which can be compensated for in the source using the phase controlling HWP.

\paragraph*{Polarisation state tomography}

Polarisation state tomography combines statistics from projection measurements to generate the density matrix of a state. Single photon rotations are applied by a QWP and HWP before a PBS. Single qubit tomography requires four measurements and two qubit tomography requires 16. Accidental counts are removed by taking each reading with and without an electronic delay. This helps reduce noise in our measurements.

\section{Entangled state fidelities} \label{supplementary:fidelities}

Photon 1 of a polarisation entangled Bell state is injected into each input of the PST waveguide array while photon 2 is preserved in polarisation maintaining fibre. Polarisation tomography is performed on the output state of each transfer and the fidelity to the state propagating through a reference straight waveguide calculated in table \ref{table:bell_state}. A characterised model of our device is constructed using quantum state tomography. The similarity between our measured output state and predicted output state is also calculated in table \ref{table:bell_state} for each transfer.

\begin{table}[h]
	\begin{tabular}{ c  c  c }
    \hline \hline
\bf{Input Waveguide}&\bf{Fidelity}&\bf{Similarity} \\ 

1 & $0.991\pm0.005$ &  $0.986\pm0.005$ \\ 
6 & $0.957\pm0.011$ &  $0.985\pm0.011$ \\
10 & $0.965\pm0.007$ & $0.990\pm0.007$ \\ \hline
    \end{tabular}
\caption{Fidelity between the output polarisation entangled state after PST and after propagation through a reference straight waveguide. Similarity is calculated between the output polarisation state and our predicted output from single photon characterisation.}
\label{table:bell_state}
\end{table}

We demonstrate PST applied to an entangled state of varying purity. Purity is controlled through applying a delay to one component of the polarisation qubit. Table \ref{table:decohered} presents the fidelity of our measured state against the same state propagation through the reference straight waveguide. Decohered PST is performed on transfer from waveguide 1 to waveguide 11.

\begin{table}[h]
	\begin{tabular}{ c  c  c }
    \hline \hline
\bf{Delay $(\mu\mathrm{m})$}&\bf{Fidelity}&\bf{Similarity} \\ 

0 & $0.964\pm0.010$ &  $0.971\pm0.010$ \\ 
50 & $0.963\pm0.012$ &  $0.981\pm0.012$ \\
100 & $0.984\pm0.005$ & $0.978\pm0.005$ \\
150 & $0.971\pm0.011$ & $0.981\pm0.011$ \\ \hline
    \end{tabular}
\caption{Fidelity between our output decohered state after PST and after propagation through a reference straight waveguide. Similarity between our measured and predicted output states.}
\label{table:decohered}
\end{table}

\end{document}